\documentstyle[aps,prl,multicol,epsf]{revtex}

\newcommand{\bq}{\begin{equation}}
\newcommand{\eq}{\end{equation}}
\newcommand{\bn}{\begin{eqnarray}}
\newcommand{\en}{\end{eqnarray}}

\begin{document}
\draft
\title{
Kondo effect and anti-ferromagnetic correlation in transport through tunneling-coupled 
double quantum dots}
\author{Bing Dong and X. L. Lei}
\address{Department of Physics, Shanghai Jiaotong University, 1954 Huashan Road, Shanghai 
200030, P. R. China}

\maketitle
\thispagestyle{empty}
\begin{abstract}
We propose to study the transport through tunneling-coupled double quantum dots (DQDs)
connected in series to leads, using the finite-$U$ slave-boson mean field approach 
developed initially by Kotliar and Ruckenstein [Phys. Rev. Lett. {\bf 57}, 1362 (1986)].
This approach treats the dot-lead coupling and the inter-dot tunnelling $t$ 
nonperturbatively at arbitrary Coulomb correlation $U$, thus allows the anti-ferromagnetic 
exchange coupling parameter $J=4t^2/U$ to appear naturally. 
We find that, with increasing the inter-dot hopping, the DQDs manifest three distinct 
physical scenarios: the Kondo singlet state of each dot with its adjacent lead, the spin 
singlet state consisting of local spins on each dot and the doubly occupied bonding 
orbital of the coupled dots.
The three states exhibit remarkably distinct behavior in transmission spectrum, linear and 
differential conductance and their magnetic-field dependence. 
Theoretical predictions agree with numerical renormalization group and Lanczos 
calculations, and some of them have been observed in recent experiments.

\end{abstract}

\begin{multicols}{2}
The prediction of the Kondo effect in a quantum dot
has been verified by observations. \cite{Goldhaber1}
Very recently, with the experimental breakthroughs,
\cite{Oosterkamp,Jeong}
much theoretical interest has been payed to
transport through tunneling-coupled double quantum dots (DQDs) to investigate the 
interplay between the Kondo effect and the anti-ferromagnetic correlation.
Aono {\it et al} \cite{Aono1,Aono2} and Georges {\it et al} \cite{Georges} employed an 
infinite-$U$ slave-boson mean-field (SBMF) theory to the problem. Aguado and Langreth 
\cite{Aguado} generalized this method to an out-of-equilibrium
situation. However, the predicted abrupt jump in
conductance at certain condition, was proved to be an artifact of the model. 
\cite{Izumida} Furthermore, the necessity of the artificial introduction of a magnetic 
coupling parameter indicates the weakness of this kind of infinite-$U$ SBMF theory in 
treating the kinetic exchange process in DQDs. \cite{Izumida}

In this letter we propose to study the transport through DQDs using the finite-$U$ SBMF 
approach, initially developed by Kotliar and Ruckenstein for other system, 
\cite{KR,DS,Dong} which enforces the correspondence between the impurity fermions and the 
auxiliary bosons to a mean-field level to release the $U=\infty$ restriction. In DQDs this 
approach allows one to treat both the dot-lead coupling and the inter-dot tunnelling $t$ 
nonperturbatively at arbitrary strength of Coulomb correlation $U$ and thus an 
anti-ferromagnetic exchange coupling parameter $J=4t^2/U$ to appear naturally. The derived 
conductance exhibits smooth change with changing tunneling strength and magnetic field 
rather than the abrupt transition showing up in the previous SBMF theory. The agreement of 
the present results with those of numerical renormalization group (NRG) and numerical 
Lanczos calculation justifies the suitability of the proposed finite-$U$ SBMF approach to 
DQDs.

Transport through tunneling-coupled double quantum dots $L$ and $R$
(assuming a single level $\epsilon_{\eta}$ in the $\eta$th dot,
$\eta=L,R$) connected separately to leads $L$ and $R$ (having energy
$\epsilon_{\eta k}$ for Bloch state $k$, $\eta=L,R$)
can be modeled by a two-impurity Anderson Hamiltonian with intra-dot
Coulomb repulsion $U_{\eta}$ and dot-lead coupling $V_{\eta}$,
and inter-dot hopping $t$. In the presence of a magnetic
field $B$ the electron energies are split into
$\epsilon_{\eta \sigma}= \epsilon_{\eta} + \sigma h_{\rm B}$ and
$\epsilon_{\eta k \sigma}= \epsilon_{\eta k} + \sigma h_{\rm B}$
for up ($\sigma=1$) and down ($\sigma=-1$) spins,
where $2h_{\rm B}=g\mu_{\rm B}B$ ($g$ is the Land\'e factor and
$\mu_{\rm B}$ the Bohr magneton).
The physical problem of DQDs described by this Hamiltonian can be dealt with
by means of the finite-$U$ slave-boson technique of
Kotliar and Ruckenstein \cite{KR,DS,Dong}
using the following effective Hamiltonian in an enlarged Hilbert space
which contains, in addition to the electron (fermion) annihilation (creation)
operators $c_{\eta k \sigma}$ ($c_{\eta k \sigma}^{\dagger}$)
in the leads and $c_{\eta \sigma}$ ($c_{\eta \sigma}^{\dagger}$)
in the dots, a set of four auxiliary boson operators
$e_{\eta}$ ($e_{\eta}^{\dagger})$, $p_{\eta\sigma}$
($p_{\eta\sigma}^{\dagger}$) and $d_{\eta}$ ($d_{\eta}^{\dagger})$
for the $\eta$th dot, which act, respectively, as projectors
onto the empty, singly occupied (with spin up and down) and
doubly occupied electron states:
\begin{eqnarray}
H_{eff}&=&\sum _{\eta, k, \sigma} \epsilon _{\eta k \sigma} c_{\eta k \sigma}^\dagger
c_{\eta k \sigma}
+ \sum _{\eta,\sigma} \epsilon _{\eta\sigma} c_{\eta\sigma}^\dagger
c_{\eta\sigma}\nonumber\\
&+&\sum_{\eta} U_{\eta}d_{\eta}^\dagger d_{\eta}+t\sum _\sigma
(z_{L\sigma}^{\dagger} c_{L\sigma}^\dagger c_{R\sigma}z_{R\sigma}
+{\rm H.c.})\nonumber\\
&+&\sum _{\eta, k,\sigma }V_{\eta}(c_{\eta k, \sigma}^\dagger
c_{\eta\sigma}z_{\eta \sigma}
+ {\rm H.c.})\nonumber\\
&+&\sum_{\eta} \lambda_\eta^{(1)} (\sum_\sigma p_{\eta\sigma}^\dagger
p_{\eta\sigma}+e_{\eta}^{\dagger} e_{\eta}+d_{\eta}^{\dagger }d_{\eta}-1)\cr
&+&\sum_{\eta,\sigma }\lambda
_{\eta \sigma }^{(2)}(c_{\eta\sigma }^{\dagger }c_{\eta\sigma }-p_{\eta\sigma
}^{\dagger }p_{\eta\sigma }-d^{\dagger }_{\eta}d_{\eta}).
\label{hamiltonian}
\end{eqnarray}
Here, the operator $z_{\eta\sigma}$ is given as
\bn
z_{\eta\sigma}&=&\left (1-d_{\eta}^\dagger d_{\eta}-p_{\eta\sigma}^\dagger p_{\eta\sigma}
\right)^{-1/2} \left ( e_{\eta}^\dagger p_{\eta \sigma}+p_{\eta\bar {\sigma}}^\dagger 
d_{\eta}\right )\cr
&\hphantom{=}&\left (1-e_{\eta}^\dagger e_{\eta}-p_{\eta\bar {\sigma}}^\dagger
p_{\eta\bar {\sigma}}\right )^{-1/2}.
\en
The constraints, i.e. the completeness relation
$\sum_{\sigma}p_{\eta\sigma}^\dagger p_{\eta\sigma}+e_{\eta}^\dagger
e_{\eta}+d_{\eta}^\dagger d_{\eta}=1$ and the condition for the
correspondence between fermions and bosons
$c_{\eta\sigma}^\dagger c_{\eta\sigma}=p_{\eta\sigma}^\dagger
p_{\eta\sigma}+d_{\eta}^\dagger d_{\eta}$, which should be imposed
for each dot, have been incorporated with Lagrange multipliers
$\lambda_{\eta}^{(1)}$ and $\lambda_{\eta \sigma}^{(2)}$ in the effective
Hamiltonian. The form (2) of $z_{\eta\sigma}$ operators is chosen
to reproduce the correct $U\rightarrow 0$ limit in the mean field
approximation.

From the effective Hamiltonian (1) one can derive equations of motion
of the slave-boson operators.
Then we use the mean-field approximation in these equations and in the
constraints and the effective Hamiltonian, in which all the boson operators
are replaced by their expectation value.
It is believed that at low temperatures, the mean field approximation is
correct for describing the Kondo-type transport through QD in the Kondo
regime. The Fermi operators in these equations are bilinear and their
expectation value can be expressed, with the help of the Langreth analytical
continuation rules for the close time-path Green's function \cite{Langreth},
by the distribution Green's functions of both dots
\bq
G_{\eta\eta \sigma}^<(\omega)=2i \frac { \tilde{\Gamma}_{\eta \sigma} f_{\eta}(\omega)
[(\omega-\tilde{\epsilon}_{\bar{\eta}\sigma})^2+\tilde{\Gamma}_{\bar{\eta}\sigma}^2]
+\tilde{\Gamma}_{\bar{\eta}\sigma} f_{\bar{\eta}}(\omega)\tilde{t}_{\sigma}^2 
}{\Delta_{\sigma}(\omega)}
\eq
($\bar{\eta}\neq {\eta}$), in which
$f_{\eta}(\omega)$ is the Fermi distribution function of $\eta$th
lead,
\bn
\Delta_{\sigma}(\omega)&=&[(\omega-\tilde{\epsilon}_{L\sigma}+
i\tilde{\Gamma}_{L\sigma})(\omega-\tilde{\epsilon}_{R\sigma}+
i\tilde{\Gamma}_{R\sigma})-\tilde{t}_{\sigma}^2]\nonumber\\
&\times&[(\omega-\tilde{\epsilon}_{L\sigma}-i\tilde{\Gamma}_{L\sigma})
(\omega-\tilde{\epsilon}_{R\sigma}-i\tilde{\Gamma}_{R\sigma})-
\tilde{t}_{\sigma}^2],
\en
$\tilde{t}_{\sigma}\equiv t|z_{L\sigma}||z_{R\sigma}|$,
$\tilde{\epsilon}_{\eta\sigma}\equiv\epsilon_{\eta\sigma}+
\lambda_{\eta\sigma}^{(2)}$, and
$\tilde{\Gamma}_{\eta \sigma}=\Gamma_{\eta}|z_{\eta\sigma}|^2$,
 $\Gamma_{\eta} =  \pi \sum_{k} |V_{\eta}|^2 \delta(\omega
 -\epsilon_{\eta k})$
being the coupling constant between the $\eta$th QD and the lead.
These 8 equations, together with the statistical average of the 6
constrains, forms the closed set of 14 equations
($\eta=L,R$ and $\sigma=1,-1$) for total fourteen unknowns
($e_{\eta}$, $p_{\eta\sigma}$, $d_{\eta}$, and $\lambda_{\eta}^{(1)}$,
$\lambda_{\eta\sigma}^{(2)}$).
The current through DQD is given by
$
I=\frac{e}{h}\int_{-\infty}^{+\infty}
T(\omega)[f_{L}(\omega)-f_{R}(\omega)]$, in which
\bq
T(\omega)=\sum_{\sigma}\tilde{\Gamma}_{L\sigma}\tilde{\Gamma}_{R\sigma}\tilde{t}_{\sigma
}^2 |\Delta_{\sigma}(\omega)|^{-2}
\eq
is the transmission probability.

This letter focuses on the situation of two identical QDs:
$\epsilon_{L}=\epsilon_{R}=\epsilon_d$, $V_{L}=V_{R}=V$ and
$\Gamma_{L}=\Gamma_{R}=\Gamma$.
In the linear limit the conductance can be written as
\bq
G=\frac{e^2}{h}\sum_{\sigma}T_{\sigma}(0)=
\frac{e^2}{h}\sum_{\sigma}\frac{4\Gamma^2t^2}{4
\Gamma^2t^2+(\Gamma^2-t^2+\frac{\tilde{\epsilon}_{d\sigma}^2}
{(\tilde{t}_{\sigma}/t)^2})^2}.
\eq
In the case of $t\ll U$, the anti-ferromagnetic exchange coupling can be
described by $J=4t^2/U$. The Kondo temperature of the QD is given
by $T_{\rm K}=U\sqrt{\beta}\exp(-\pi/\beta)/2\pi$ 
[$\beta=-2U\Gamma/\epsilon_{d}(U+\epsilon_{d})$], where the energy level of the QD, 
$\epsilon_d$, is measured from the Fermi level of adjacent lead.
Hereafter in the text and in the figures we will always use $\Gamma$ as the
energy units.  

Fig.\,1 shows the linear conductance $G$ and the total occupied number $N$
for $t=1,2$ and $8$ at $U=14$, as functions of $\epsilon_d$ in the absence of magnetic 
field. Depending on the hopping strength $t$, three different
physical scenarios can be seen. (1) In the case of $t\lesssim 1$, when inter-dot coupling 
is weaker than the dot-lead coupling ($J/T_{\rm K} \lesssim 0.57$) $G$ has a single peak 
in the $N\simeq 2$ regime ($-0.75<\epsilon_d/U<-0.25$), indicating that Kondo singlet 
states consisting of each individual dot and adjacent lead dominate. The inter-dot 
tunneling serves mainly as the conducting bridge across the two dots. The peak conductance 
increases with increasing $t$ and reaches the unitary value $2e^2/h$ at $t=1$. (2) In the 
case of $t=2$, when inter-dot coupling becomes
stronger than the dot-lead coupling ($J/T_{\rm K}= 2.3$), $G$ descends slightly in the 
$N\simeq 2$ range, exhibiting a shallow valley. The increasing inter-dot tunneling 
together with the strong intra-dot Coulomb repulsion favors the formation of a spin 
singlet state consisting of local spins on each dot. (3) In the case of $t=8$, when the 
inter-dot correlation is strong enough to compensate the intra-dot Coulomb repulsion 
($t\gtrsim U/2$), the doubly occupied bonding (even) orbital of the coupled dots are 
favorite. $G$ exhibits a double-peak structure with a deep valley almost down to zero, and 
the Kondo feature disappears completely.

Fig.\,2 shows the inter-dot-hopping-dependent conductance. One can see the continuous 
appearance of the three different scenarios when the inter-dot hopping
changes. $G$ increases with increasing $t$ from zero, reaches the maximum at around
$t\simeq 1$ and then decreases.
The single peak structure of $G$ at $U=8$ turns into a pattern having a shoulder
for larger $U$, and finally the shoulder becomes a monotonously growing curve at 
$U\rightarrow \infty$ limit.
These results are in fairly good agreement with those of
numerical Lanczos calculation \cite{Busser} and NRG calculation \cite{Izumida}
for finite Coulomb interaction $U$.

In the present model the energies 
$\epsilon_{e(o)\sigma}=\tilde{\epsilon}_{d\sigma}\mp
\tilde{t}_{\sigma}$ can be considered as the energy levels of the even and odd states
(in the case without the magnetic field they are independent of the spin and we will omit 
the $\sigma$ index). The energy separation between the odd and even
states $\epsilon_{o}-\epsilon_{e}=2\tilde {t}$ equals $2t$ for $U=0$   
and is largely suppressed by the intra-dot Coulomb repulsion.
On the other hand, in the Kondo-dominanted regime ($J/T_{\rm K}\lesssim 1$, $N\simeq 2$) 
the resonant energy level $\tilde{\epsilon}_d$ locates very near to the chemical
potential of the lead, such that the ratio  
$P\equiv\tilde{\epsilon}_{d}^2/(\tilde{t}/t)^2$ is much smaller than 1 [Fig.\,2(b)], and 
the conductance (6) is simplified to $G=\frac{2e^2}{h}\frac{4t^2} {(1+t^2)^2}$, which 
reaches its maximum value at $t=1$. 

The physical origins of the transport
properties can be understood in terms of the transmission spectra $T(\omega)$, which are 
depicted in Figs.\,3(a)-(c) for $\epsilon_d = -2$ and $U=8,14,20$ in the absence of 
voltage bias. $T(\omega)$
reaches the maximum $1$ at $\omega=\epsilon_{e}$
and $\epsilon_{o}$. In the case of $t=1$ where the intra-dot Kondo correlation
is predominant over the inter-dot coupling ($J/T_K=0.78, 0.57,
0.43$ for $U=8,14$ and 20, respectively), $\epsilon_e$ and $\epsilon_o$ are closed 
together that $T(\omega)$ exhibits only a single peak. 
In the case of $t=2$ where the inter-dot coupling grows over the intra-dot Kondo 
correlation ($J/T_K=3.1, 2.3, 1.7$ for $U=8,14$ and 20, respectively)
two peaks in $T(\omega)$ show up [Fig.\,3(b)]. An example of strong inter-dot coupling 
($t=8$) is plot in Fig.\,3(c), where two peaks in $T(\omega)$ are so greatly separated 
that the valley bottom is nearly opaque. For $\epsilon_d =-2$ the zero-frequency line is  
in between $\epsilon_{e}$ and $\epsilon_{o}$ but always closer to the even peak.
 
A more significant and experimentally measured quantity is the differential
conductance $dI/dV$, which have been calculated at finite source-drain
biases by assuming a symmetric voltage drop, $\mu_L=-\mu_R=eV/2$.
The calculated results of $dI/dV$ are plotted in Fig.\,3(d)-(f) as a
function of $eV/T_{\rm K}$. The differential conductance behavior changes
continuously with increasing $t$, but exhibits remarkably different characteristics for 
the above three typical scenarios. The predicted double peak of the Kondo resonance in the 
$dI/dV$-$V$ curves has been observed in a recent measurement (in Ref.\cite{Jeong}). The 
sharp dips with negative differential conductivity showing up in the theoretical curves, 
may be smeared out by the dephasing processes at finite bias voltages and other complexity 
in a real experiment. We look forward to further subtle experiment to test it.

Fig.\,4 shows the linear conductance in the presence of external magnetic
fields (MC), as a function of Zeeman energy $h_{\rm B}$ (a) and as a function of 
$\epsilon_d/U$ (b). Again, $t=1$ and $t=2$ represent two distinct physical situations.
This behavior of the MC can be understood from the magneto-transmission spectra
as shown in Figs.\,5(a)-(c).
In the case of $t=1$, for which the $T(\omega)$ exhibits a single
broad maximum around $\omega=0$ at $h_{\rm B}=0$, the transmission probability, and
thus the linear conductance $G$,
decline with increasing $h_{\rm B}$, indicating
a negative MC and a weakening of the Kondo effect [Fig.\,5(a)].
The stronger the intra-dot Coulomb repulsion, the faster
the conductance $G$ decreases with increasing $h_{\rm B}$.
When $t\gtrsim 2$, both the even peak and odd peak in the
transmission spectrum $T(\omega)$ in the absence of the magnetic field
may split into two peaks corresponding to up and down spins
for strong magnetic fields, giving rise to a four-peak structure
[Fig.\,5(c) for $h_{\rm B}=0.4$].
Note that in the presence of intra-dot Coulomb repulsion
the splitting between the up and down
peaks is smaller than the Zeeman energy $2h_{\rm B}$
(similar to the situation of the single QD \cite{Goldhaber1,Dong})
and always wider in the even channel than in the odd channel.

In the case of $t=2$, where the even and odd peaks are moderately separated
and the valley in $T(\omega)$ locates not far from $\omega=0$ for
$h_{\rm B}=0$, the up-spin peak of the even channel may emerge
with the down-spin peak of the odd channel at suitable magnetic fields,
giving rise to a triple-peak pattern. As a consequence,
$T(\omega)$ at $\omega=0$ increases with increasing magnetic field up to
a certain strength, yielding a positive MC and an enhancement of the Kondo
effect. Of course, further splitting of the up and down spin peaks by
the continuing increase of the magnetic field will lead to rapid
drop of the transmission probability (and the conductance $G$) after
reaching the maximum. As can be seen from Fig.\,4(a) the position of the
conductance maximum, is strongly dependent on the Coulomb interaction $U$.
It should be noted that, this rapid drop, is apparently smooth rather
than abrupt.
A similar behavior of MC has been found by Aono {\it et. al.} \cite{Aono2}
in their $U\rightarrow \infty$ model, but their prediction that
MC suddenly becomes zero at a certain magnetic field, seems to be an artifact of
the model.

Finally, we suggest that the magneto-transmission probability can be observed
in experiment by measuring the $dI/dV$-$V$ characteristic in the presence
of magnetic fields. Figs.\,5(d)-(f) demonstrate that $dI/dV$ transits
from single zero-bias peak to multiple-peak pattern depending on the
inter-dot hopping $t$.

In conclusion we have studied the transport through DQDs using the finite-$U$ SBMF 
approach. Due to competition between the Kondo and anti-ferromagnetic correlations, the 
DQDs manifest three distinct physical scenarios as the inter-dot hopping $t$ increases 
from zero: the Kondo singlet state of each dot with its adjacent lead, the spin singlet 
state consisting of local spins on each dot, and the doubly occupied bonding orbital of 
the coupled dots. Although transition between them is continuous, the three states exhibit 
remarkably distinct behavior in transmission spectrum, linear and differential conductance 
and their magnetic-field dependence. These features have been partly observed recently, 
and further experiments are needed for a full test of the theory.    

This work was supported by the National Natural Science Foundation
of China (grant Nos. 60076011 and 90103027), the Special Funds for Major
State Basic Research Project (grant No. 2000683), and the
Shanghai Municipal Commission of Science and Technology.

\vspace{2cm}

\begin{center}
{\bf Figure Captions}
\end{center}

\vspace{0.2cm}
\noindent
{\bf Fig.1} Linear conductance $G$ (thick lines) and total charge number $N$ (thin lines) 
vs. the QD energy level $\epsilon_d$ at $U=14$ for inter-dot hopping $t=1$ (solid line), 
$2$ (dashed line), and $8$ (dotted line) repectively. 

\vspace{0.2cm}
\noindent
{\bf Fig.2} (a) Inter-dot hopping-dependent linear conductance through DQDs with 
$\epsilon_d=-2$ at different Coulomb interaction $U=8$, $14$, $20$, and $\infty$. (b) The 
parameter $P$ vs. the inter-dot hopping $t$ for $U=14$ and different energy levels 
$\epsilon_d=0$, $-1$, $-2$, $-4$, $-5$ and $-6$.

\vspace{0.2cm}
\noindent
{\bf Fig.3} Transmission spectrum [(a)-(c)] and $dI/dV$-$V$ curves [(d)-(f)] for 
$\epsilon_d=-2$ and different values of the hopping $t=1$, $t=2$, and $t=8$.

\vspace{0.2cm}
\noindent
{\bf Fig.4} (a) The magneto-conductance vs. magnetic field $h_{\rm B}$ for 
$\epsilon_d=-2$. (b) The magneto-conductance vs. the QD energy level for 
$h_{\rm B}=0$, $0.1$, and $0.2$. 

\vspace{0.2cm}
\noindent
{\bf Fig.5} Magneto-transmission spectrum [(a)-(c)] and the $dI/dV$-$V$ curves [(d)-(f)] 
for $U=8$, $\epsilon_d=-2$ at different magnetic fields. 

\end{multicols}
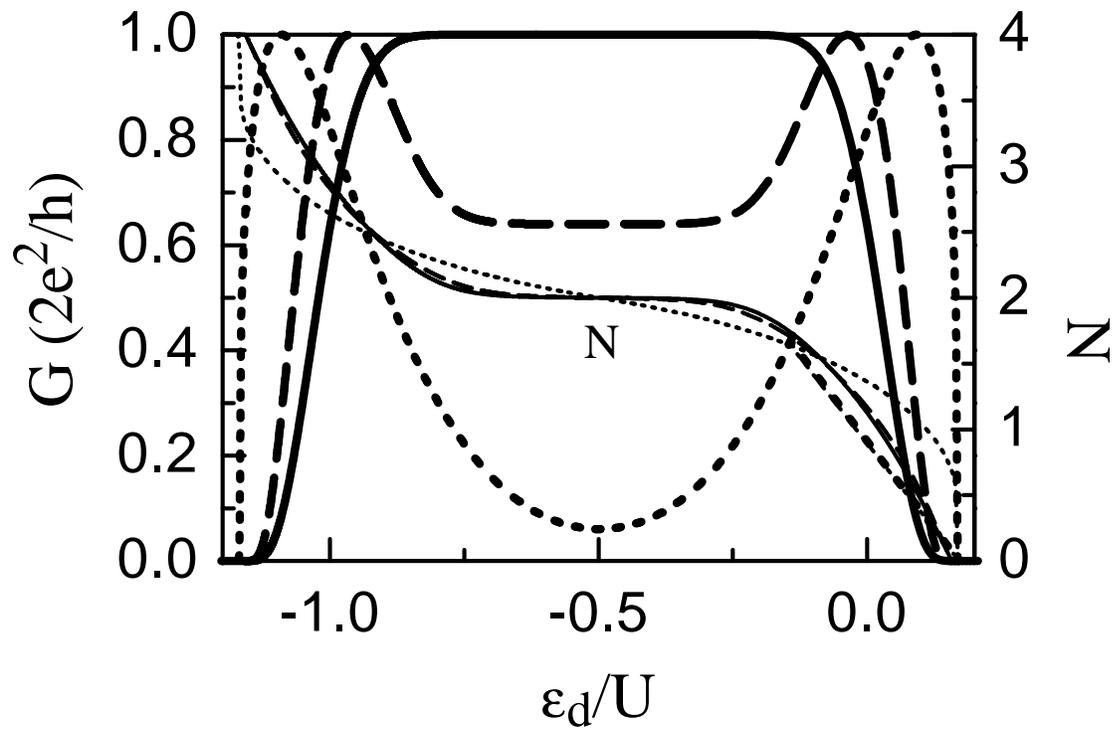
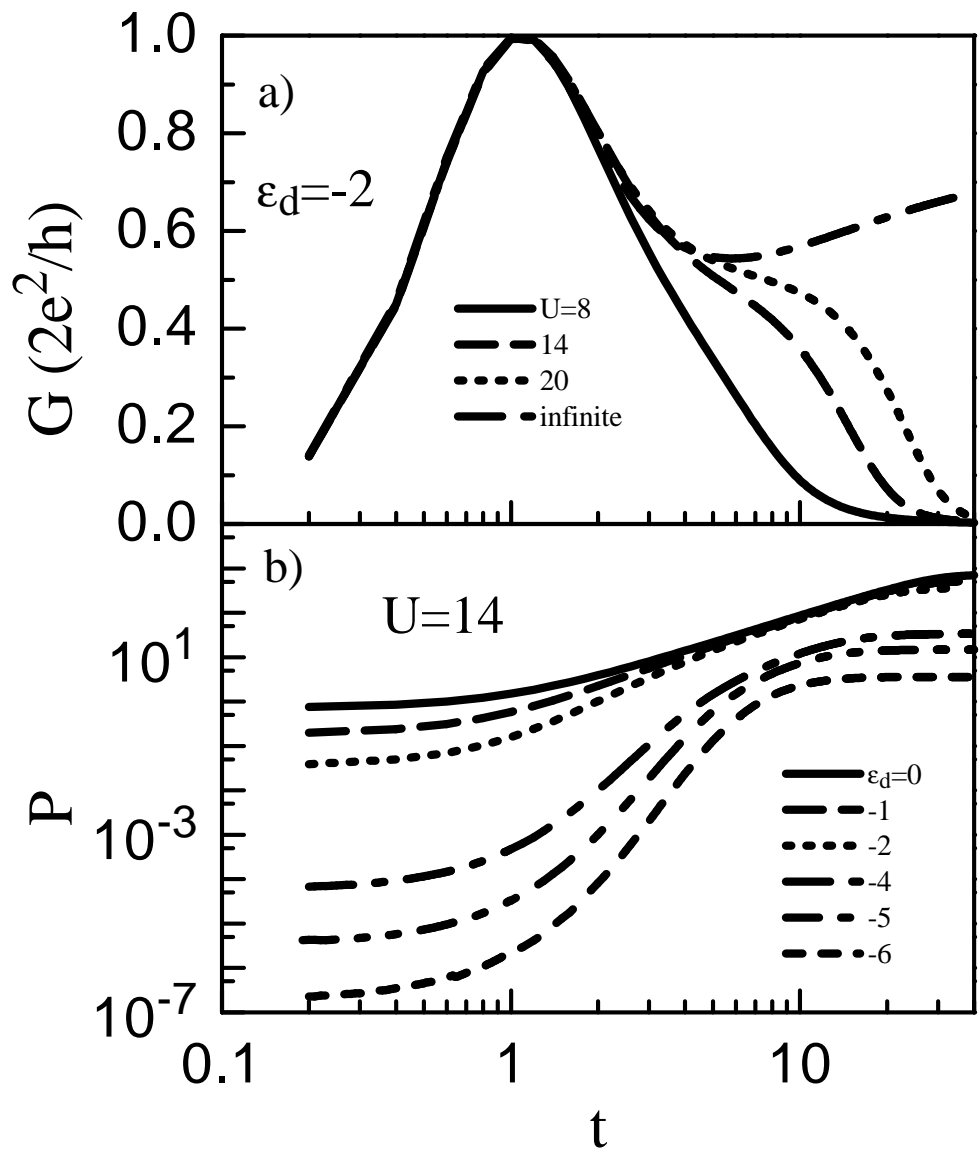
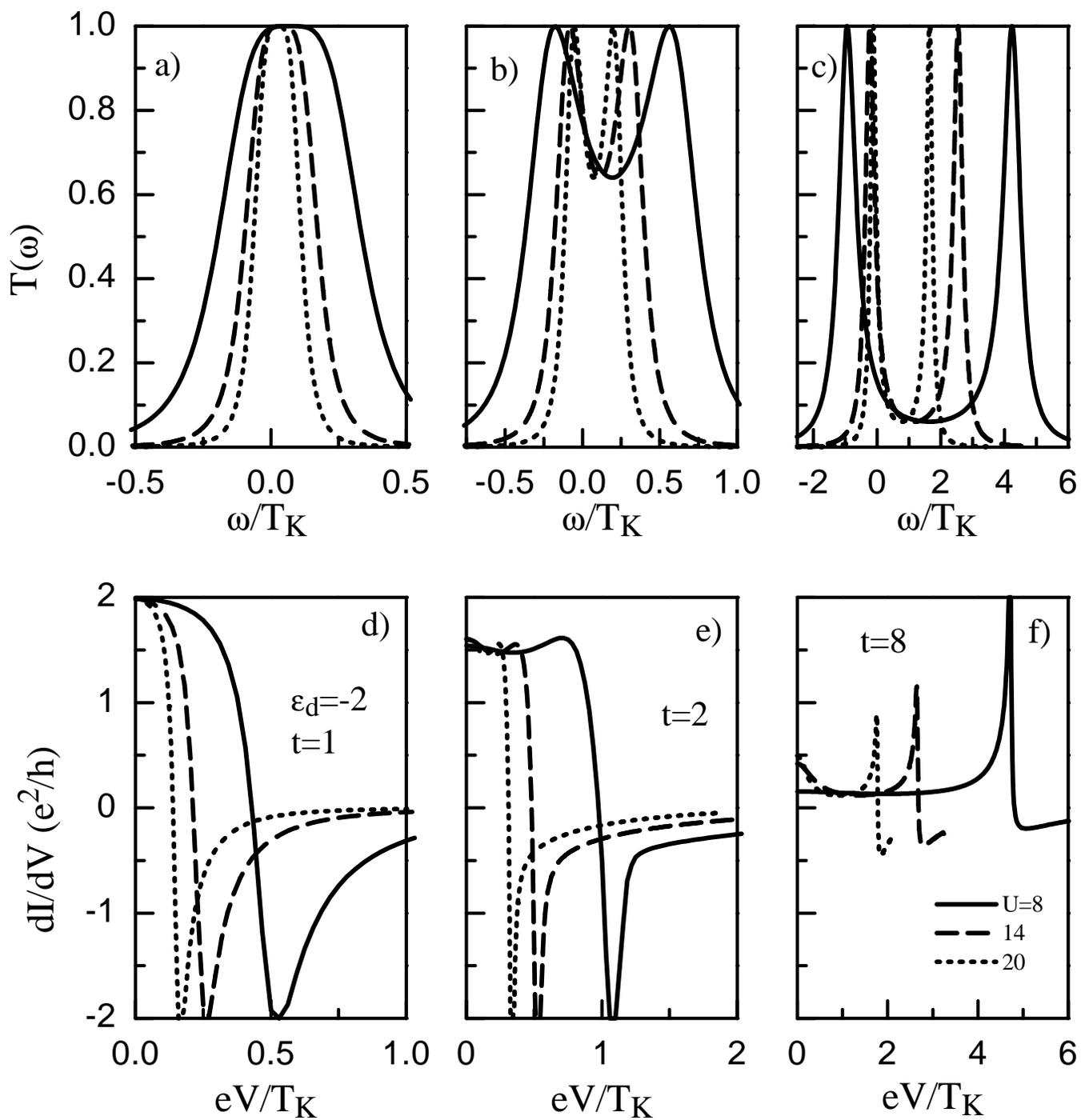
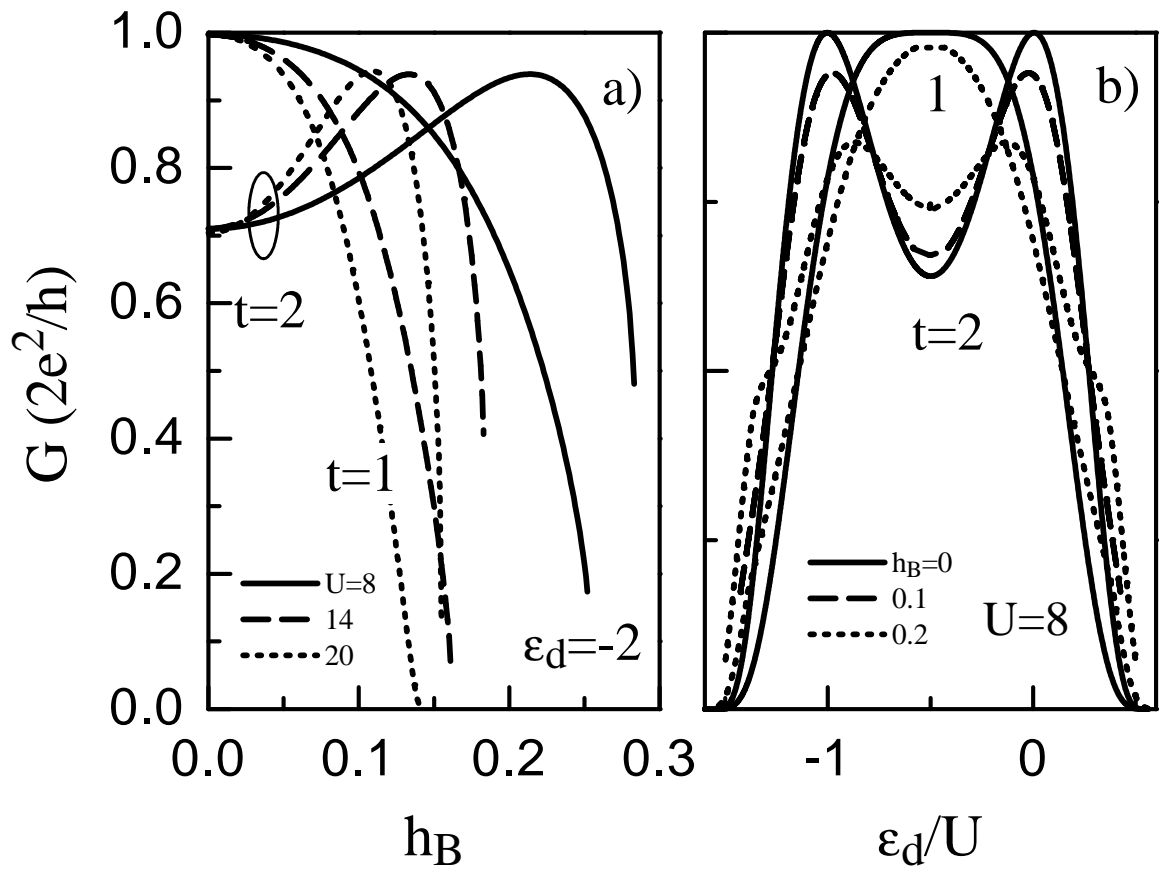
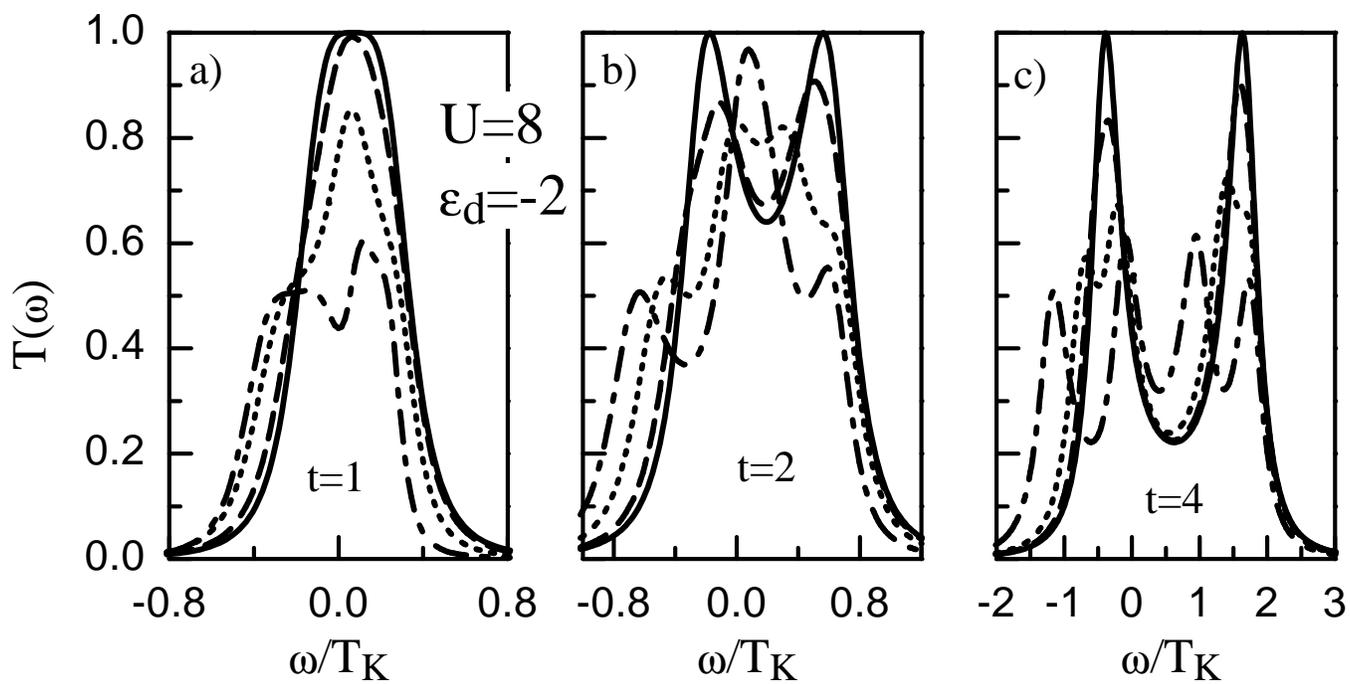
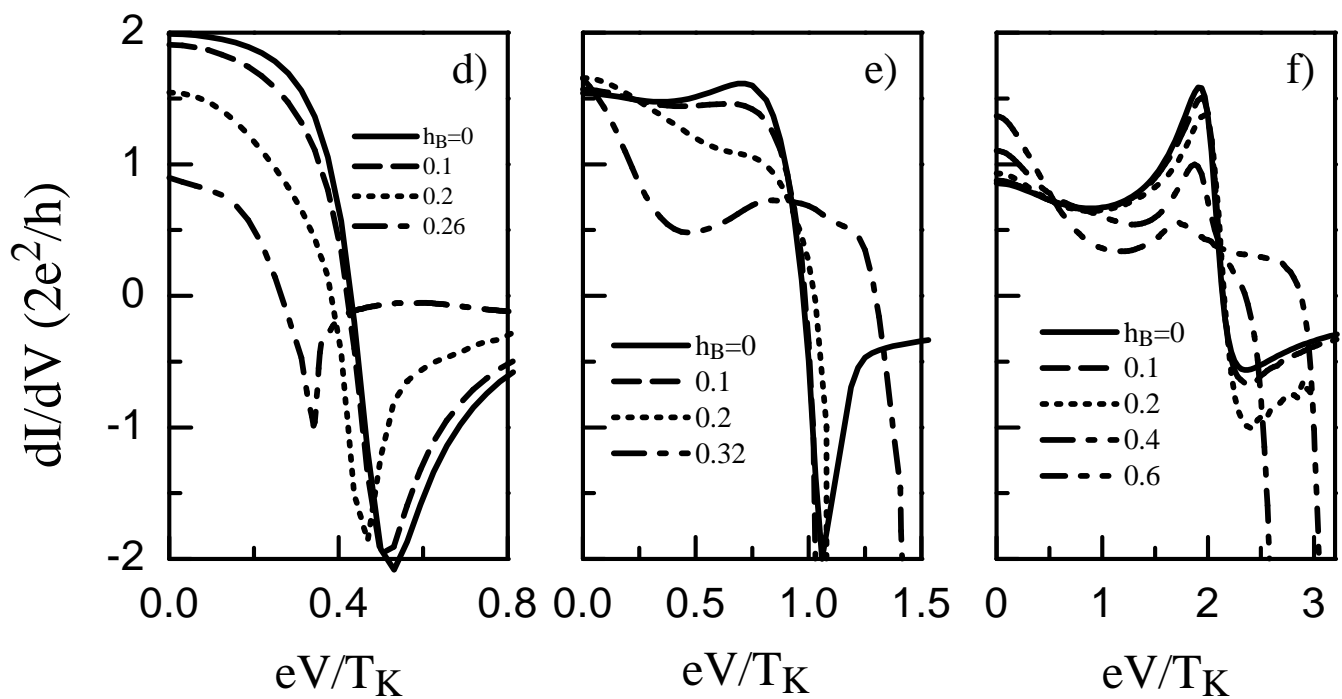
\end{document}